\documentclass[12pt]{aipproc}
\layoutstyle {6x9}
\usepackage{epsfig}
\SetInternalRegister\hbadness{8000}
\newcommand{\be}{\begin{equation}}
\newcommand{\ee}{\end{equation}}

\newcommand\doingARLO[2][]{%
  \ifx\mmref\undefined #1\else #2\fi
  }

\begin{document}
\title{Generalized Weinberg's approach to the $a_0/f_0$ case}

\author{Yu. S. Kalashnikova}{
  address={ITEP, 117218, B.Cheremushkinskaya 25, Moscow, Russia},
  email={yulia@heron.itep.ru}
}
\begin{abstract}
The problem of whether it is possible to distinguish composite from 
elementary particles is studied in the framework of Weinberg's approach.
The possibility to extend this approach to the case of unstable particles 
in the presence of inelastic channels is considered. The interplay 
between the 
low-energy scattering data and the admixture of a bare state in the 
resonance is discussed, and the implications for the $a_0(980)/f_0(980)$ 
case are outlined.  
\end{abstract}
\date{}
\maketitle

The nature of $a_0(980)/f_0(980)$-mesons remains the most enigmatic 
question of meson spectroscopy. Quark models \cite{Isgur} predict the 
$1^3P_0$ $q \bar q$ states made of light quarks to exist at about 1 GeV.
However, the vicinity of $K \bar K$ threshold suggests that significant 
$\frac{1}{\sqrt{2}}(u \bar u \pm d \bar d)s \bar s$ admixture should be 
present in the wave functions of these mesons, either in the form of 
compact four-quark states \cite{Achasov} or in the form of $K \bar K$ 
molecules \cite{WeiIs}, \cite{Markushin}. The related question 
to be addressed in this regard is of the relative role of $s$- and 
$t$-channel force in the formation of such molecules \cite{IsSp}.

The approaches of \cite{WeiIs}, \cite{Jue} and \cite{Markushin} conclude 
that the $t$-channel force can be dominant in producing the attraction 
in $K \bar K$ channel necessary to form the $a_0$ and $f_0$ as hadronic 
molecules. On the other hand, as $a_0$/$f_0$ couple strongly to $K \bar 
K$ channel, one expects drastic unitarity effects, which are responsible 
for the dressing of the bare states seeded into mesonic continuum, the 
phenomenon described in the framework of coupled channel models 
\cite{Markushin}, \cite{port}, \cite{Tornqvist}.

Therefore the observed features of the $\pi\pi$, $\pi\eta$ and $K \bar 
K$ spectra could be explained both by potential-type interaction in 
these systems and by existence of ``bare'' confined states strongly 
coupled to mesonic channels, and the question persists whether it is
possible to distinguish between different assignments for $a_0$/$f_0$. 

Many years ago S.Weinberg \cite{SWein} has considered a similar problem 
of "elementarity" of the deuteron, expressing the effective range 
$n$-$p$ parameters in terms of field renormalization constant $Z$, which 
defines the admixture of a bare elementary-particle state in the 
deuteron. It was shown that the low-energy $n$-$p$ data are consistent 
with small value of $Z$, so that the deuteron is indeed a 
molecular-type particle made of proton and neutron.

To apply this approach, three requirements are needed. The particle 
should couple to a two-body channel with the threshold close 
to the nominal mass; this two-body channel should have zero orbital 
momentum; the particle must be stable, otherwise the factor $Z$ is not 
defined. First two requirements are met in the $a_0/f_0$ case, while the 
third one is not met: the decays $f_0 \rightarrow \pi\pi$ and 
$a_0 \rightarrow \pi\eta$ are known to be the main source of the 
width for these mesons. Nevertheless, it appears to be possible \cite{af}
to generalize Weinberg's approach to the case of unstable 
particles. 

The starting point of such generalization is the dynamical scheme of the 
coupled channel model. It is assumed that the hadronic state is 
represented symbolically as
\be
|\Psi \rangle = \left(\sum_{\alpha}c_{\alpha}|\psi _{\alpha}\rangle\atop 
\sum_i
\chi_i |M_1(i)M_2(i) \rangle \right),
\label{state}
\ee
where the index $\alpha$ labels bare confined states
$|\psi_{\alpha}\rangle$ with the probability amplitude $c_{\alpha}$, and
$\chi_i$ is the wave function in the $i$-th two-meson channel 
$|M_1(i)M_2(i)\rangle$. The wave function $|\Psi\rangle$ obeys the 
equation
\be
\hat {\mathcal H} |\Psi\rangle = E |\Psi\rangle,~~
\hat {\mathcal H} = \left(\begin{array}{cc}
\hat{H}_c&\hat{V}\\
\hat{V}&\hat{H}_{MM}
\end{array}
\right),
\label{ham}
\ee
where $\hat {H}_c$ defines the discrete spectrum of bare states,
$\hat {H}_c |\psi_{\alpha}\rangle = E_{\alpha} |\psi_{\alpha}\rangle$,
$\hat{H}_{MM}$ includes the free-meson part as well as direct
meson-meson interaction (e.g., due to $t$- or $u$-channel exchange 
forces),
and the term $\hat {V}$ is responsible for dressing the bare states.
The latter is specified by the transition form factor
$f_{M_1(i)M_2(i)}^{\alpha}(p)$,
\be
\langle \psi_{\alpha}|\hat {V}|M_1(i)M_2(i)\rangle = 
f_{M_1(i)M_2(i)M}^{\alpha}(p),
\label{ff}
\ee
where $p$ is the relative momentum in the mesonic system 
$M_1(i)M_2(i)$. The function $f$ decreases with $p$ with some range 
$\beta$ whose scale is set by the size scale of hadronic wave functions; 
the estimate for $\beta$ is to be of order of a few hundred MeV.

In a simple case of only one bare state $|\psi_0 \rangle$
and only one hadronic channel ($|K \bar K \rangle $) the system 
of equations (\ref{ham}) is easily solved, yielding for the $K
\bar K$ scattering amplitude the form
\be
F_{K \bar
K}(k,k;E)=-\frac{2\pi^2mf_{KK}^2(k)}{E-E_0+g_K(E)},~~k=\sqrt{mE},
\label{amp}
\ee
where
\be
g_K(E)=\int \frac{f_{KK}^2(p)}{\frac{p^2}{m}-E-i0} d^3p.
\label{g}
\ee
If the system possesses a bound state with the energy $-\epsilon$, the 
admixture of a bare state in the bound state wave function, 
$|c_0|^2=\cos^2\theta$, is defined from the expression
\be 
\tan^2\theta=\int
\frac{f_{KK}^2(p)d^3p}{(\frac{p^2}{m}+\epsilon)^2}.
\label{angle}
\ee
In the small binding limit $\sqrt{m\epsilon}
\ll \beta$ it is possible to express the effective range
parameters in terms of the binding energy $\epsilon$ and angle $\theta$
in a model-independent way (for the details see \cite{af}).  
The relations between scattering length $a$ and effective 
range $r_e$, and the binding energy $\epsilon$ and $Z=\cos^2\theta$ read
\be
a=\frac{2(1-Z)}{2-Z}R +O(1/\beta),~~r_e=-\frac{Z}{1-Z}R 
+O(1/\beta),~~R=1/\sqrt{m\epsilon},
\label{wlr}
\ee
coinciding with the ones obtained in \cite{SWein}.

In the case of unbound state one is to consider the continuum counterpart 
of $Z$, the spectral density of the bare state introduced in 
\cite {bhm} and given by the expression
\be
w(E)=2\pi m k |c_0(E)|^2,
\label{w}
\ee
where $c_0(E)$ is found from the system of Eqs. (\ref{ham}) in the 
continuum. Due to the normalization condition
\be
\int_0^{\infty}w(E)dE=1-Z~~ {\rm or} ~~1,
\label{intw}
\ee
depending on whether there is a bound state or not, all the information 
on $Z$ is encoded in the $w(E)$ too, and the generalization to the 
multichannel case is straightforward. 

If one is interested only in the phenomena near $K \bar K$ threshold, 
it appears possible to express the spectral density $w(E)$ in terms of  
hadronic observables. Indeed, one can make use of the smooth 
dependence on energy of the integral $g_P$ for the light 
pseudoscalar channel, similar to (\ref{g}). 
Then the near-threshold $K \bar K$ scattering amplitude is 
given by the Flatt{\`e}-type expression
\be
F_{K \bar K}= -\frac{1}{2k} \frac{\Gamma_K}{E-E_f+i\frac{\Gamma_K}
{2}+i\frac{\Gamma_P}{2}},
\label{ampin}
\ee
where
$$
E_f=E_0-\bar E_K -\bar E_P,~~\Gamma_K
=\bar g_{K \bar K}\sqrt{mE},~~\bar g_{K \bar K}=4\pi^2mf_{K0}^2,
$$
$$
\bar E_K = 4\pi m\int_0^\infty f_{KK}^2(p)dp,~~f_{K0}=f_{KK}(0),
$$
and $\bar E_P$ and $\frac{1}{2}\Gamma_P$ are the real and imaginary 
parts of the integral $g_P$ averaged over $K \bar K$ near-threshold 
region.

The spectral density can be written out as
\be
w(E)=\frac{1}{2\pi}\frac{\Gamma_P+
\bar g_{K \bar K}\sqrt{mE}\Theta(E)}
{(E-E_f-\frac{1}{2}\bar g_{K \bar
K}\sqrt{-mE}\Theta(-E))^2+
\frac{1}{4}(\Gamma_P+\bar g_{K \bar K}\sqrt{mE}\Theta(E))^2}.
\label{wfl}
\ee
Eq. (\ref{wfl}) expresses the spectral density $w(E)$ in terms
of hadronic observables (Flatt{\`e} parameters),
just in the same way as Weinberg's
factor $Z$ is expressed in terms of hadronic observables (effective 
range
parameters) via Eqs. (\ref{wlr}). Thus, Eq.
(\ref{wfl}) generalizes Weinberg's result to the case of
unstable particles.

It is clear from the expression (\ref{wfl}) that it is the singularity 
structure of the scattering amplitude which governs the behaviour of 
spectral density. In the elastic case in the presence of a 
bound state the pole positions in the $k$ plane are given by
\be
k_1=i\sqrt{m\epsilon}, ~~k_2=-i\sqrt{m\epsilon}\frac{2-Z}{Z}.
\label{wroots}
\ee
For a deuteron-like situation, i.e.
for $Z \ll 1$, the second pole is far from the threshold and even moves to
infinity in the limit $Z \rightarrow 0$. On the other hand, if $Z$ is
close to one, i.e. if there is considerable admixture of an elementary
state in the wave function of the bound state, both poles are 
near threshold. In the limiting case $Z \rightarrow 1$ the poles 
are located equidistantly from the point $k=0$. With inelasticity included,
one expects the spectral density to be enhanced in the 
vicinity of the amplitude poles. If the poles are located in the
near-threshold region, the spectral density in this region would be 
large. If, on the contrary, there is only one near-threshold
pole, a considerable part of the spectral density is smeared over a much
wider energy interval, which is a signal that the
bare state admixture in the near-threshold resonance is small.
So there is a one-to-one correspondence between
the value of $Z$ (or the behaviour of $w(E)$) and the
pole counting scheme suggested by Morgan \cite{morgan}. 

Several Flatt{\`e}-like fits to $\pi\eta$ and $\pi\pi$ spectra were 
analysed in \cite{af}, and the pole positions and near-threshold spectral 
densities for these fits were found. The results are given in Table 1 
together with the integral
\be
W_{a_0 (f_0)}=\int^{50 MeV}_{-50 MeV} w_{a_0 (f_0)}(E)dE \ .
\label{W}
\ee
of the spectral density over the region containing the $K \bar K$ 
threshold. The limit of integration is chosen to be twice as large as the 
peak width of the $a_0/f_0$ mesons. As the function $w(E)$ is normalized 
to unity, the integral (\ref{W}) is a direct measure of bare state 
admixture in the $a_0/f_0$ mesons. 

\begin{table}[t]
\begin{tabular}{c|c|c|c||c|c|c|c}
Ref.&$k_1$&$k_2$&$W_{a_0}$&Ref.&$k_1$&$k_2$&$W_{f_0}$\\
\hline
\cite{Teige}&-104+i55&104-i111&0.49&\cite{SND}&-58+i107&58-i729&0.17\\
\cite{Bugg}&-134+i71&134-i199&0.29&\cite{CMD2}&-65+i97&65-i477&0.23\\
\cite{AchKi}&-129+i44&129-i250&0.24&\cite{KLOE}&-69+i100&69-i804&0.14\\
\cite{AchKi}&-126+i73&126-i212&0.29&\cite{AchGu}&-84+i17&84-i351&0.21\\
\cite{KLOE}&-102+i97&102-i199&0.36&&&&\\
\end{tabular}
\caption{Pole positions (in MeV/c) and $W$ for various fits to $a_0$ 
(left) and $f_0$ (right) mesons.}
\end{table}

The spectral densities for various Flatt{\`e} fits are shown at Fig.1. The 
$a_0$-meson looks like an above-threshold
phenomenon, with considerable part of bare state spectral density peaked
near $K \bar K$ threshold. The $f_0$ is a below-threshold
resonance, with some small part of spectral density peaked below
threshold. Obviously, this part can be viewed as Weinberg's "$Z$", smeared
due to the presence of inelasticity, and this "$Z$" is definitely small.

The interrelation between pole positions and near-threshold fraction of 
the bare state spectral is clearly seen from the Table 1. The case of  
a pair of pole singularities near the threshold corresponds to the bare 
state accidentally seeded into near-threshold region; the admixture of a 
bare state should be large in this case. In the opposite case of small
bare state admixture one has only one stable pole position near threshold.
 
Indeed, in the $a_0$ case the fits lead to more equidistant positions of 
poles than in the $f_0$ case, and the near-threshold fraction of $w(E)$ 
is more sizable for the $a_0$ meson. Still, even for $a_0$-meson it is, 
averagely, about 30\%, so the $a_0(980)$ contains a large admixture of 
mesonic components. As for the $f_0$ meson, there is only one 
near-threshold pole, the near-threshold 
fraction of $w(E)$ is about 20\% or less, and mesonic 
component in $f_0$ is large.

We conclude, in such a way, that the simple $q \bar q$ assignment is
inadequate for the $f_0(980)$-meson, and is not very appropriate for the
$a_0(980)$-meson.

\begin{figure}[t]
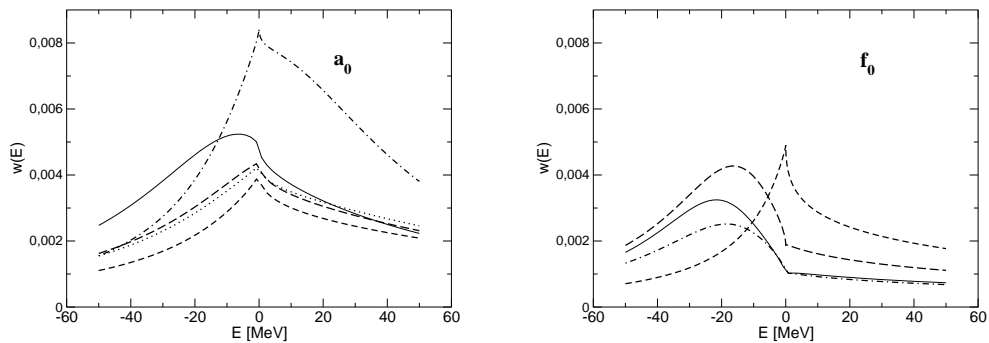

\centerline{\epsfxsize=6cm\epsfbox{wa0_1.eps}\hspace*{1cm}\epsfxsize=6cm\epsfbox{wf0_1.eps}}
\caption{a) Spectral densities $w(E)$ for the $a_0$ meson
based on the Flatt{\`e} parameters taken from
Ref. \cite{Teige} (dashed-dotted line), Ref. \cite{Bugg} (dotted line),
Ref. \cite{AchKi} (dashed line), Ref. \cite{AchKi} (long dashed line), 
and
Ref. \cite{KLOE} (solid line).
b) Spectral densities $w(E)$ for the $f_0$ meson
based on the Flatt{\`e} parameters taken from
Ref. \cite{SND} (solid line), Ref. \cite{CMD2} (long-dashed line),
Ref. \cite{KLOE} (dashed-dotted line), and Ref. \cite{AchGu} (dotted
line).}
\end{figure}

The author would like to thank V. Baru, J. Haidenbauer, C. Hanhart and 
A. Kudryavtsev for fruitful cooperation.
Financial support of the grants 02-02-04001/DFG-436 RUS/113/652 and 
NSh-1774.2003.2 is gratefully acknowledged. This work is supported by 
the Federal Programme of the Russian Ministry of Industry, Science and 
Technology No 40.052.1.1.1112.

\bibliography{sample}

\begin{thebibliography}{99}
\bibitem{Isgur} S. Godfrey, and N. Isgur, {\em Phys. Rev.} {\bf D32}, 189 
(1985).
\bibitem{Achasov} R. L. Jaffe, {\em Phys. Rev.} {\bf D15}, 267, 281 (1977);
N. N. Achasov, S. A. Devyanin, G. N. Shestakov, {\em Phys. Lett.} {\bf B96}, 
168, (1980). 
\bibitem{WeiIs} J. Weinstein and N. Isgur, {\em Phys. Rev.} {\bf D27}, 588 
(1979).
\bibitem{Markushin} M. P. Locher, V. E. Markushin, and H. Q. Zheng, 
{\em Eur. Phys. J.} {\bf C4}, 317 (1998).
\bibitem{IsSp} N. Isgur and J. Speth, {\em Phys. Rev. Lett.}, {\bf 77}, 2332 
(1996).
\bibitem{Jue} D. Lohse, J. W. Durso, K. Holinde, and J. Speth, 
{\em Phys. Lett.} {\bf B234}, 235 (1990); G. Janssen, B. C. Pearce, K. Holinde, 
and J. Speth, {\em Phys. Rev.} {\bf D52}, 2690 (1995).
\bibitem{port} E. van Beveren, C. Dullemond, and G. Rupp, {\em Phys. Rev.}
{\bf D21}, 772 (1980); G. Rupp, E. van Beveren, and
M. D. Scadron, {\em Phys. Rev.} {\bf D65}, 078501 (2002); 
E. van Beveren and G. Rupp, {\em hep-ph/0304105}.
\bibitem{Tornqvist} N. A. T\"ornqvist, {\em Z. Phys.} {\bf C68}, 674 (1995); 
N. A. T\"ornqvist and M. Roos, {\em Phys. Rev. Lett.} {\bf 76}, 1575 (1996).
\bibitem{SWein} S. Weinberg, {\em Phys. Rev.} {\bf 137} B672 (1965).
\bibitem{af} V. Baru, C. Hanhart, J. Haidenbauer, Yu. Kalashnikova, 
and A. Kudryavtsev, {\em hep-ph/0308129}.
\bibitem{bhm} L. N. Bogdanova, G. M. Hale, and V. E. Markushin, {\em Phys. 
Rev.} {\bf C44}, 1289 (1991).
\bibitem{morgan} D. Morgan, {\em Nucl. Phys.} {\bf A543}, 632 (1992).
\bibitem{Teige} S. Teige et al., {\em Phys. Rev.} {\bf D59}, 012001 (2001).
\bibitem{Bugg} D. V. Bugg, V. V. Anisovich, A. Sarantsev, and B. S. Zou, 
{\em Phys. Rev.} {\bf D50}, 4412 (1994).
\bibitem{AchKi} N. N. Achasov and A. N. Kiselev, {\em Phys. Rev.} {\bf D68},
014006 (2003).
\bibitem{KLOE} A. Antonelli, eConf. C020620 THAT06 (2002).
\bibitem{SND} M. N. Achasov et al., {\em Phys. Lett.} {\bf B485}, 349 (2000).
\bibitem{CMD2} R. R. Akhmetshin et al., {\em Phys. Lett.} {\bf B462}, 380 
(1999).
\bibitem{AchGu} N. N. Achasov and V. V. Gubin,  {\em Phys. Rev.} {\bf D63}, 
094007 (2001).
\end{thebibliography}

\end{document}